 \documentclass[preprint]{aastex}

\shortauthors{Moon \& Eikenberry}
\shorttitle{QPOS IN THE LARGE X-RAY FLARES FROM LMC X--4}

\begin{document}


%\title{Collapsed Cores in Globular Clusters, \\
%    Gauge-Boson Couplings, and AAS\TeX\ Examples}
\title{LARGE X-RAY FLARES FROM LMC X--4: \\
DISCOVERY OF MILLI-HERTZ QUASI-PERIODIC  \\
OSCILLATIONS AND QPO-MODULATED PULSATIONS}


%\author{S. Djorgovski\altaffilmark{1,2,3} and Ivan R. King\altaffilmark{1}}
%\affil{Astronomy Department, University of California,
%    Berkeley, CA 94720}

%\author{C. D. Biemesderfer\altaffilmark{4,5}}
%\affil{National Optical Astronomy Observatories, Tucson, AZ 85719}
%\email{aastex-help@aas.org}

%\and

%\author{R. J. Hanisch\altaffilmark{5}}
%\affil{Space Telescope Science Institute, Baltimore, MD 21218}


%\altaffiltext{1}{Visiting Astronomer, Cerro Tololo Inter-American Observatory.
%CTIO is operated by AURA, Inc.\ under contract to the National Science
%\altaffiltext{2}{Society of Fellows, Harvard University.}
%\altaffiltext{3}{present address: Center for Astrophysics,
%    60 Garden Street, Cambridge, MA 02138}
%\altaffiltext{4}{Visiting Programmer, Space Telescope Science Institute}
%\altaffiltext{5}{Patron, Alonso's Bar and Grill}

%\author{Dae-Sik Moon and Stephen S. Eikenberry}
\author{DAE-SIK MOON and STEPHEN S. EIKENBERRY}
\affil{Department of Astronomy, Cornell University, Ithaca, NY 14853 \\
moon,eiken@astrosun.tn.cornell.edu}


\begin{abstract}
We report the discovery of milli-hertz (mHz) quasi-periodic oscillations (QPOs) 
and QPO-modulated pulsations during
large X-ray flares from the high-mass X-ray binary pulsar LMC X--4 
using data from the {\it Rossi X-Ray Timing Explorer} (RXTE).
The lightcurves of flares show that,
in addition to $\sim$74 mHz coherent pulsations,
there exist two more time-varying temporal structures at frequencies of
$\sim$0.65--1.35 and $\sim$2--20 mHz.
These relatively long-term structures appear in the 
power density spectra  as 
mHz QPOs and as well-developed sidebands around the
coherent pulse frequency as well,
indicating that the amplitudes of the coherent pulsation is
modulated by those of the mHz QPOs.
One interesting feature is that,
while the first flare shows symmetric sidebands around the coherent pulse frequency,
the second flare shows significant excess emission in the lower-frequency
sidebands due to the $\sim$2--20 mHz QPOs.
We discuss the origin of the QPOs using a combination of
the beat-frequency model and a modified version of the Keplerian-frequency model.
According to our discussion,
it seems to be possible to attribute the origin of the $\sim$0.65--1.35 
and $\sim$2--20 mHz QPOs 
to the beating between the rotational frequency
of the neutron star and the Keplerian frequency of 
large accreting clumps near the corotation radius and
to the orbital motion of clumps at Keplerian radii of 2$-$10 $\times$ 10$^9$ cm, respectively.
\end{abstract}


\keywords{accretion, accretion disks --- pulsars: individual (LMC X--4) --- stars: neutron --- X-rays: stars}


\section{Introduction}

Since the first discovery of an X-ray pulsar in the Cen X$-$3 system,
it has been generally accepted that accretion disk plays a key role
in the evolution of accretion-powered X-ray binary pulsars (AXBPs).
The instantaneous luminosity and spin rates are mainly determined
by accretion rates from the surrounding disk
through the release of gravitational potential energy and 
the transfer of angular momentum.
If the precessing accretion disk is tilted with respect to the orbital plane of the binary
and periodically lies in a plane containing the line-of-sight to the pulsar,
the long-term super-orbital motions detected from a few AXBPs
can also be explained.
In spite of its critical importance and extensive previous studies, however,
the detailed properties of the accretion disk around AXBP are still poorly known.

One useful phenomenon for studying the accretion disk around AXBP is 
quasi-periodic oscillations (QPOs)
which have been considered to be related to the motion of the accretion disk, 
especially the innermost part.
About 10 AXBPs have been detected with X-ray QPOs 
with frequency of 5--220 mHz to date 
(Boroson et al. 2000, and references therein).
According to the beat-frequency model (BFM),
which has been the canonical model for QPOs in X-ray binaries,
the observed QPO frequency ($\nu_{\rm QPO}$) is a beat frequency
between the coherent spin frequency of the pulsar ($\nu_{\rm s}$)
and the Keplerian frequency ($\nu_{\rm K}$) of the innermost disk,
$\nu_{\rm QPO}$ = $\nu_{\rm K}$ $-$ $\nu_{\rm s}$,
at the magnetosphere boundary of the pulsar \cite{as85,let85}.
Another model often used in explaining the X-ray binary QPOs
is the Keplerian-frequency model (KFM),
which interprets the QPOs as a result of absorption of X-rays by
inhomogeneities in the inner edge of the accretion disk, and, subsequently, 
the observed QPO frequency as the frequency of the Keplerian motion, 
$\nu_{\rm QPO}$ = $\nu_{\rm K}$ \cite{vdket87}.
Since the KFM requires $\nu_{\rm s}$ $<$ $\nu_{\rm QPO}$ = $\nu_{\rm K}$
(otherwise the accretion could be prohibited by the so-called 
``propeller effect" [Illarinov \& Sunyaev 1975]), 
which is not always true for the case of AXBPs,
a modification is needed to explain the observed frequencies of QPOs from AXBPs 
(e.g., Kommers, Chakrabarty, \& Lewin 1998, hereafter KCL98).
Hence, while the BFM explains QPOs as luminosity oscillations, 
the KFM explains QPO as beaming oscillations.

LMC X--4 is a persistent, disk-fed, high-mass AXBP in the Large Magellanic Cloud
with a pulsational period of $\sim$13.5 s.
The average X-ray (2$-$25 keV) luminosity is 
$\sim$2 $\times$ 10$^{38}$ ergs s$^{-1}$
and the optical companion is a 14th magnitude O-type star of $\sim$15 $M_{\odot}$.
The $\sim$1.4 day orbital period was found by several different studies including
observations of periodic X-ray eclipses and optical brightness variations.
The long-term intensity of LMC X--4 varies with a period of $\sim$30.3 days,
indicating the possible existence of a precessing accretion disk
tilted relative to the orbital plane.
Heindl et al. (1999) pointed out that,
if this long-term variation is due to reprocessing of the X-ray beam by a disk,
Thomson scattering by a hot ionized disk must be the dominant mechanism of reprocessing.
Heemskerk \& van Paradijs (1989) estimated the average X-ray albedo of the accretion 
disk to be $\sim$0.93 from analysis of the optical light curve.

LMC X--4 has exhibited large X-ray flares which uniquely distinguish it from other AXBPs.
According to the results of Levine, Rappaport, \& Zojcheski (2000, hereafter LRZ00), 
the X-ray (2$-$25 keV) luminosity increases up to $\sim$10$^{39}$ ergs s$^{-1}$
during the flares and the modulation factor of the pulsations
increases substantially. 
Individual pulsations are clearly visible from the raw data of the flares.
Two relatively long-term temporal structures of the flares were identified
with characteristic time scales of 
$\sim$150 and  $>$ 400 s, respectively.
The period of the flaring episodes is not well estimated,
but is roughly equal to the orbital period.
Although it is very likely that the flaring episodes are related to 
increased rates of accretion,
no proposed mechanism successfully explains the origin of the flares (e.g., LRZ00).

\section{Observations and Analysis}

RXTE detected several large flares from LMC X--4
on 1996 August 19 during $\sim$36-hour continuous observations of the source.
We transformed the photon arrival times from the Good Xenon data of 
the Proportional Counter Array
to the solar system barycenter using the JPL DE400 ephemeris.
Four data segments, each with $\sim$4.5-ks length,
containing the flaring episode were obtained.

%\placefigure{fig1}

Light curves of the first and second data segments, 
obtained with the all five Proportional Counter Units in the 2$-$5 keV range,
are shown in Fig. 1 with a time resolution of 4 s. Several large flares are 
clearly identified, and we shall call the flares of the first data segment ``FLARE 1" and 
the second data segment ``FLARE 2."
The phase of the $\sim$35-d super-orbital motion,
based on the results of Ilovaisky et al. (1984), for the flares is 0.66 $\pm$ 0.23,
while the $\sim$1.4-d binary orbital phases, based on the ephemeris of LRZ00,
for FLARE 1 and FLARE 2 are $\sim$0.26 and $\sim$0.30, respectively.
The difference between the start time of data segment 1 and 
data segment 2 is $\sim$6000 s. 
Three different temporal structures during the flares are evident in Fig. 1:
(1) strong burst-like structures with $\sim$700--1500 s period
(we shall call these ``big QPOs" [BQPOs])
(2) relatively weak oscillations with $\sim$50--500 s period
(we shall call these ``medium QPOs" [MQPOs]), 
and (3) $\sim$13.5-s coherent pulsations
in the small window of Figure 1a.

%\placefigure{fig2}

Fig. 2 shows the Leahy-normalized power density spectra (PDS) 
of FLARE 1 and FLARE 2 obtained with a time resolution of 2$^{-6}$ s.
The main windows show the PDS around the coherent pulse frequency, $\sim$0.074 Hz.
The small windows on the left sides show the low-frequency (mHz) PDS
in the logarithmic frequency scale, while the small windows on the right sides
show the detailed structures of the sidebands around the coherent frequency.
At least three characteristics are evident in the PDS of Fig. 2: 
(1) a strong peak at the coherent pulse frequency, 
(2) well-developed significant sidebands around the coherent frequency, 
and (3) several strong components at mHz frequencies.
The low-frequency powers due to the BQPOs and MQPOs 
are identified in in the left panels.
The frequency of the BQPOs decreases from $\sim$1.35 to $\sim$0.65 mHz
from FLARE 1 to FLARE 2, which is evident in Fig. 1.
The MQPOs in Fig. 1 are the several strong excess 
power features at $\sim$2$-$20 mHz.
The sidebands of FLARE 1 are very close to a symmetric distribution 
with respect to the coherent frequency, while the sidebands of FLARE 2 show
significant asymmetry between the intensities of 
the lower-frequency (LF) and the higher-frequency (HF) sidebands,
with the LF $\sim$2 times stronger than the HF sidebands.
Table~\ref{tbl-1} lists fractional root-mean-square (FRMS) amplitudes,
obtained by the methods described by van der Klis (1989),
of the QPOs and coherent pulsations in Figure 2.
Fig. 3 shows the same PDS as Fig. 2, but with the energy range of 5--13 keV.
We see almost identical structures to the 2--5 keV range.

%\placetable{table1}
%\placefigure{fig3}

As in Fig. 2, the structure of the sidebands around the coherent pulse frequency
closely resembles the structure of the low-frequency powers in the mHz frequency range.
In order to effectively compare the sideband structures with the power
distribution of the mHz QPOs,
the powers of the mHz QPOs were overlaid onto those of the sidebands: 
First the PDS of the mHz QPOs were shifted to the frequency of the
HF sidebands, and then a mirror image, with respect to the coherent frequency,
of the shifted PDS onto the frequencies of the LF sidebands was made.
Finally, the intensities of the shifted PDS of the mHz QPOs were scaled
to match those of the sidebands.
Fig. 4 compares the PDS of the sidebands (solid histogram) with the shifted, scaled
PDS (crosses) of the mHz QPOs in FLARE 1 (Fig. 4a) and FLARE 2 (Fig. 4b),  
respectively.
They show almost identical distributions 
(except for the excess emission of the LF sidebands of FLARE 2),
indicating that the amplitudes of the coherent pulsations is modulated by those
of the mHz QPOs.

%\placefigure{fig4}

\section{Discussion}

QPO-related sidebands around a coherent pulse frequency were previously detected
from the low-mass X-ray binary pulsar 4U 1626$-$67 (KCL98),
where $\sim$48 mHz QPOs modulate $\sim$130 mHz coherent pulsations.
In the low-energy PDS of this source,
the LF sidebands are significantly stronger than their HF complements.
KCL98 suggested that a structure (or ``blob")
orbiting at the Keplerian radius of the QPO frequency
produces (1) symmetric sidebands by crossing the line-of-sight once per orbit and
thus attenuating the pulsar beam
and (2) excess emission in the LF sidebands of the low-energy PDS
by reprocessing of the pulsar beam
at the beat frequencies between the pulsational frequency and the QPO
when the structure is not in the line-of-sight.
The model of KCL98 explains the observed PDS of 4U 1626$-$67,
although the nature of the orbiting structure and the reprocessing mechanism are not clear.
Since the direct comparison between the PDS of the QPOs and those of the sidebands
are not given for this source, it is not clear whether the coherent pulsations are
modulated exactly by the amplitude of the QPOs.

LMC X--4 shows at least two apparently different features from 4U 1626$-$67.
First, the mHz QPOs only appear during the large X-ray flares,
while the $\sim$48 mHz QPOs of 4U 1626$-$67 seem to have existed for more than a decade.
Next, only the MQPOs of FLARE 2 show the significant excess emission of the
LF sidebands, and do so in both the low- and high-energy bands.

We discuss the origin of the mHz QPOs from LMC X--4 via a combination 
of the BFM and a modified version of the KFM.
First, the BQPOs can be explained by the BFM as follows: 
Suppose that the observed X-ray flares are induced by 
big clumps accreting onto the surface of the neutron star at the
inner edge of the accretion disk. The beating between the coherent pulsar frequency
and the Keplerian frequency of the clumps can result in beat frequencies at the
observed QPO frequencies, $\nu_{\rm QPO}$ = $\nu_{\rm K}$ -- $\nu_{\rm s}$,
implying that the Keplerian radius corresponding to the BQPOs 
lies very close to, but inside of, 
the corotation radius for accretion to occur (Bildsten et al. 1997). 
This is consistent with the results of previous observations
\cite{net85, wet96, let00} which predict that the magnetospheric radius
of LMC X--4 is comparable to the corotation radius, 
i.e., LMC X--4 is in spin equilibrium.
The expected corotation radius for an 1.4 $M_{\odot}$ neutron star is
$\sim$ 9.7 $\times$ 10$^8$ cm.
Next, the MQPOs can be explained by a modified version of the KFM as follows:
Suppose that the observed frequencies represent the Keplerian orbital
frequencies of clumps (or ``blobs"),
implying Keplerian orbital radii of $\sim$2$-$10 $\times$ 10$^{9}$ cm.
The observed sidebands of the MQPOs, then, can be explained similarly to the 
model of KCL98, where the excess emission of the LF sidebands
is generated by reprocessing on the clump
when the clump is not in the observer's line-of-sight to the pulsar.
The model of KCL98, however, explains the excess emission of the LF
sidebands only in the soft-energy band, which is not the case for LMC X--4.
This difference can be explained by the difference in reprocessing mechanim
of accretion disks around the two sources. 
If the reprocessing mechanims on the accretion disk around
LMC X--4 is less energy sensitive, such as reflection and scattering, the excess
emission in the LF sidebands of LMC X--4 in both soft- and hard-energy bands
can be accounted for.
This argument is indeed consistent with the results of 
Heemskerk \& van Paradijs (1989),
where a very high X-ray albedo (0.93) of the accretion disk around LMC X--4 was obtained.
Heindl et al. (1999) also suggested the possibility that scattering
is the dominant reprocessing mechanism of the accretion disk around LMC X--4.
In conclusion, it seems to be likely that the BQPOs originate near the corotation radius,
while the MQPOs originate from the orbital motions of clumps 
in the accretion disk outside the corotation radius.

In Fig. 2 and Fig. 3, the excess emission of the LF sidebands appears only in FLARE 2. 
This may indicate the existence of the excess free electrons 
produced by the intense radiation from FLARE 1.
These excess free electrons then may effectively enhance the scattering 
efficiency for the radiation from FLARE 2.
On the other hand, the apparent highest frequency of the MQPOs increases
from $\sim$10 to $\sim$20 mHz in Fig 2. and Figure 3. 
Since higher frequency implies a smaller Keplerian radius,
this increase may indicate
a radial infalling motion in the accretion disk during the large X-ray flares.
If this is true, the inferred average velocity is $\sim$2 km s$^{-1}$.
If the BQPOs are due to the large accreting clumps as explained above,
it is evident that they are unique feature caused by the large X-ray flares.
During the normal state (i.e., non-flaring state),
the power at the frequencies of the MQPOs are significantly less than the power
expected from the FRMS amplitude of the large X-ray flares,
suggesting the possibility that the MQPOs are also unique features of the flaring episodes.

Although the combination of the BFM and a modified version of
the KFM seems to account for the observed temporal 
properties of the large X-ray flares, this does not rule out, at this point,
the possibility of different mechanisms for the origin of the mHz QPOs.
For example, Titarchuk \& Osherovich (2000) proposed global oscillations
in the accretion disk for the origin of QPOs.
More observations,
especially long-term monitoring of the flare activities,
are needed to investigate these scenarios more thoroughly.

\section{Conclusions}

We summarize our results as follows: 
\begin{itemize}
\item
We have discovered mHz QPOs and QPO-amplitude-modulated coherent
pulsations in large X-ray flares from LMC X--4.
\item
The aperiodic variabilities of the flares can be classified as BQPOs ($\sim$0.65--1.35 mHz)
and MQPOs ($\sim$2--20 mHz).
\item
The BQPOs can be explained by the beating between the coherent pulse
frequency of the neutron star and the Keplerian frequency of large accreting clumps
near the corotation radius
\item
The MQPOs can be explained by attenuation and reprocessing from the orbital motions of
clumps in the accretion disk outside the corotation radius.
\item
It is very likely that reflection and scattering are the dominant reprocessing mechanisms
on the accretion disk around LMC X--4, and the increase of the highest QPO frequencies
may indicate an infalling motion of the accretion disk inner edge as the flares continue. 
\item
The results of this paper present some of best and most extreme examples of coupling
between periodic and aperiodic variabilities of AXBPs.
\end{itemize}

\acknowledgments
We would like to thank the referee, Bram Boroson, for comments and suggestions.
DSM acknowledges Wynn Ho for his comments and careful reading of manuscript.
This research has made use of data obtained from the {\it
High Energy Astrophysics Science Archive Research Center}
(HEASARC), provided by NASA's Goddard Space Flight Center.
DSM is supported by NSF grant AST-9986898.
SE is supported in part by an NSF Faculty Early Careeer Development
(CAREER) award (NSF-9983830).

\clearpage
\begin{deluxetable}{lcll}
\tablecolumns{4}
\tablewidth{0pt}
\tablecaption{Fractional Root-Mean-Square Amplitude of LMC X--4\tablenotemark{a} \label{tbl-1}}
\tablehead{
\colhead{Name} & \colhead{Frequency (mHz)}     & \multicolumn{2}{c}{FRMS (\%)}  \\  
\colhead{} &  \colhead{}  &  \colhead{FLARE 1}  &  \colhead{FLARE 2} \\ }
\startdata
BQPOs & 0.5$-$1.8   &    22 $\pm$ 8   &  60 $\pm$ 22 \\
MQPOs & $\;$2$-$15  &    30 $\pm$ 4   &  53 $\pm$ 7 \\
Pulsations\tablenotemark{b} & 60$-$90  &    18 $\pm$ 2   &  28 $\pm$ 2 \\
\label{table1}
\enddata
\tablenotetext{a}{Energy range is 2--5 keV.}
\tablenotetext{b}{including the sidebands}
\end{deluxetable}

\clearpage
\begin{figure}
\plotone{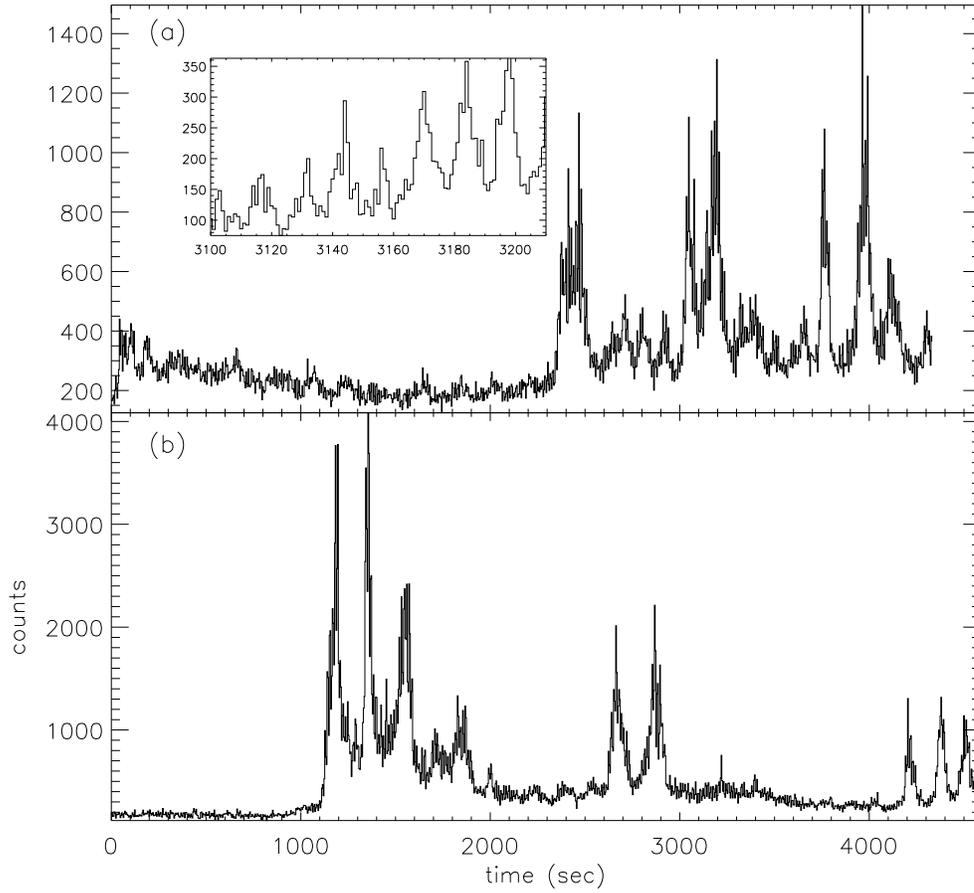}
\caption{Lightcurves, in the 2--5 keV range with 4-s time resolution,
of data segment 1 (a) and data segment 2 (b). 
The small window of the top picture shows individual pulsations during flares.  
The difference between the start time of data segment 1 and 2 is $\sim$6000 s.
\label{fig1}}
\end{figure}

\clearpage
\begin{figure}
\plotone{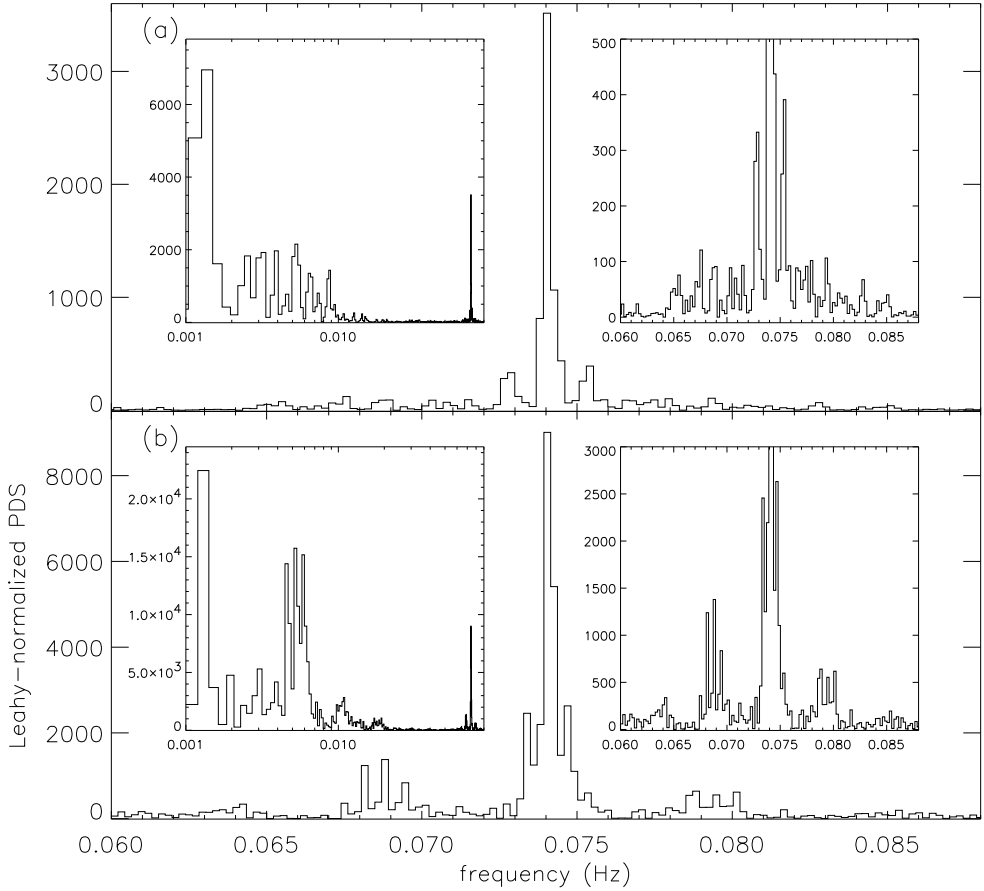}
\caption{Power-density spectra (PDS) of the lightcurves in Figure 1: 
(a) 1$\rm ^{st}$ data segment; (b) 2$\rm ^{nd}$ data segment.
The main windows show PDS around the coherent pulse frequency.
The small windows on the left sides show the low-frequency (mHz) PDS
in the logarithmic frequency scale, while the small windows on the right sides
show the detailed structures of the sidebands around the coherent frequency.
\label{fig2}}
\end{figure}

\clearpage
\begin{figure}
\plotone{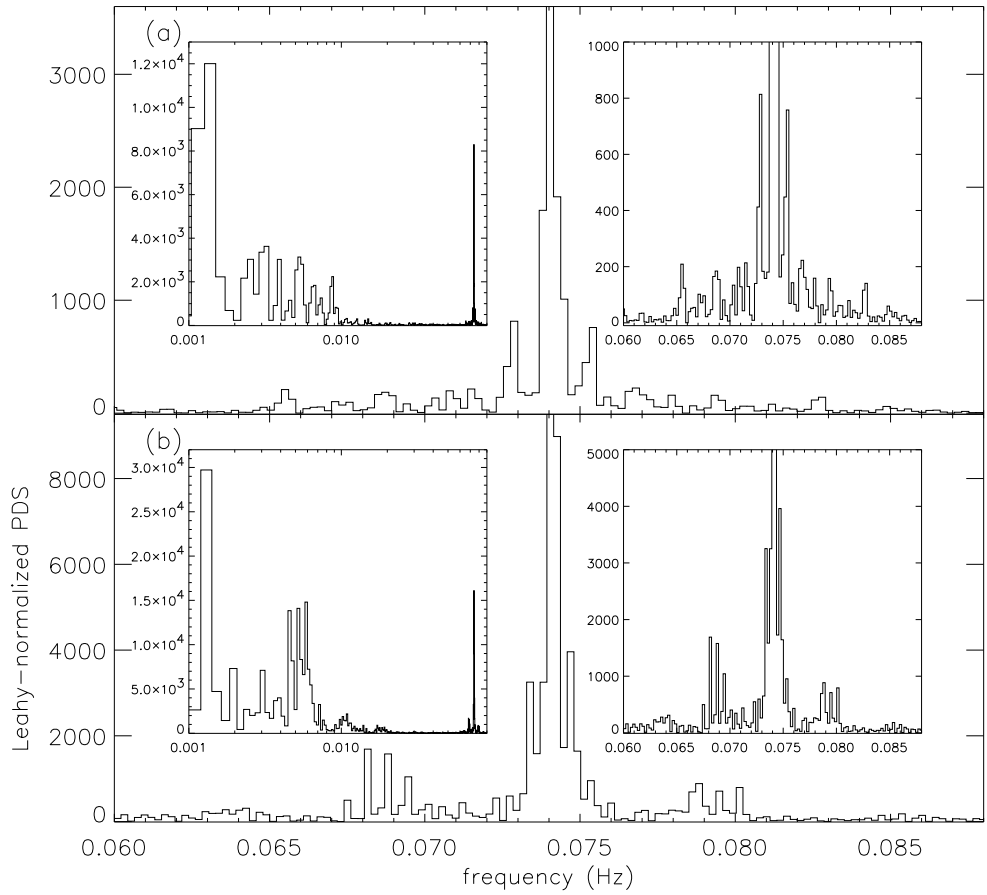}
\caption{Same as Fig. 2, but in the 5--13 keV range.
\label{fig3}}
\end{figure}

\clearpage
\begin{figure}
\plotone{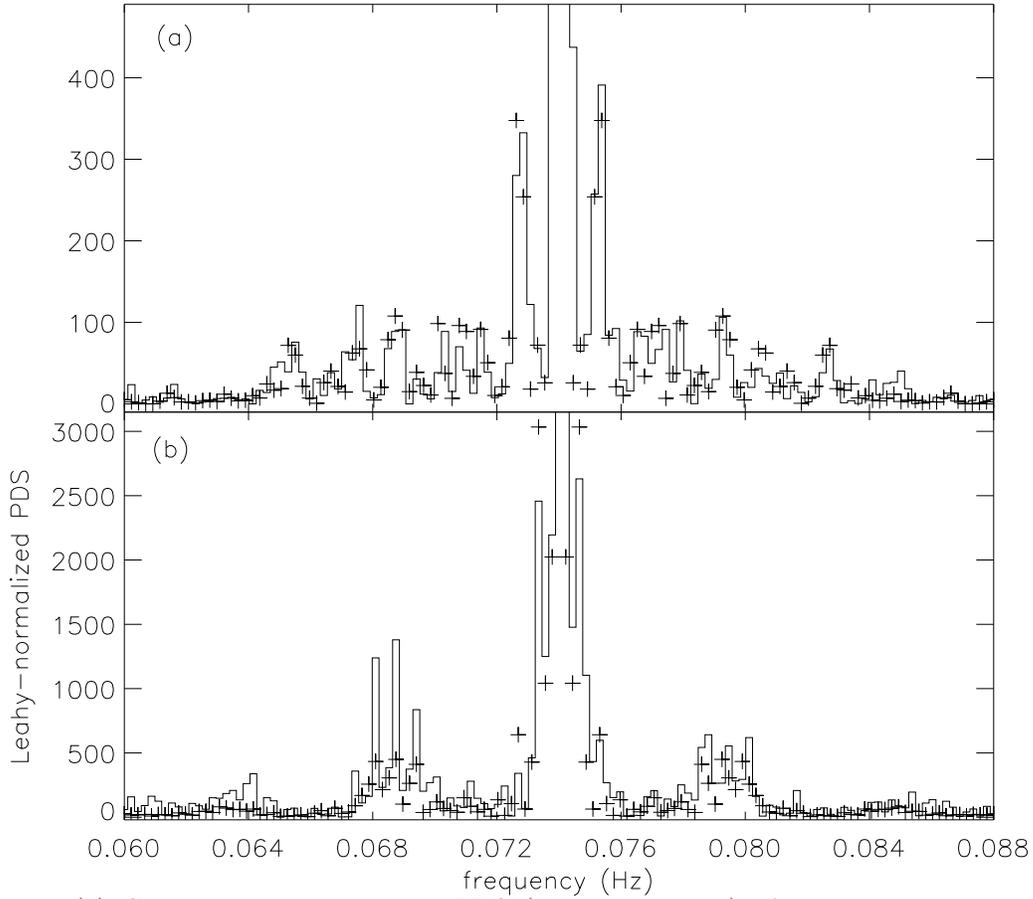}
\caption{(a) Comparison between the PDS (solid histogram)
of the sidebands and the shifted, scaled PDS (crosses) of the mHz
QPOs from data segment 1 in the 2--5 keV range.
(b) Same as (a), but for data segment 2.
Note that the lower-frequency sidebands have 
consistently larger amplitudes than their higher-freqency.
\label{fig4}}
\end{figure}

%% Tables may also be prepared as separate files. See the accompanying
%% sample file table.tex for an example of an external table file.
%% To include an external file in your main document, use the \input
%% command. Uncomment the line below to include table.tex in this
%% sample file.

%% The following command ends your manuscript. LaTeX will ignore any text
\end{document}